# A preliminary study of agility in business and production – Cases of early-stage hardware startups

Anh Nguyen-Duc
University of Southeastern Norway
Italy
angu@usn.no

Xiaofang Weng
University of Bolzano
Italy
xiaofeng.wang@unibz.it

Pekka Abrahamsson
University of Jyväskylä,
Finland
pekka.abrahamsson@jyu.fi

## ABSTRACT

[Context]Advancement in technologies, popularity of small-batch manufacturing and the recent trend of investing in hardware startups are among the factors leading to the rise of hardware startups nowadays. It is essential for hardware startups to be not only agile to develop their business but also efficient to develop the right products. [Objective] We investigate how hardware startups achieve agility when developing their products in early stages. [Methods] A qualitative research is conducted with data from 20 hardware startups. [Result] Preliminary results show that agile development is known to hardware entrepreneurs, however it is adopted limitedly. We also found tactics in four domains (1) strategy, (2) personnel, (3) artifact and (4) resource that enable hardware startups agile in their early stage business and product development. [Conclusions] Agile methodologies should be adopted with the consideration of specific features of hardware development, such as up-front design and vendor dependencies.

## CCS CONCEPTS

## 1 INTRODUCTION

The startup landscape includes not only pure software products, such as web platforms, mobile apps and desktop applications, but also products that are composed of both software and hardware units. With the Industry 4.0 revolution [1], there is an increasing amount of hardware-related products in the domain of Internet-of-things (IoT), cyber-physical systems and advanced robots. The entry threshold for starting a business around hardware-related products has never been lower due to the popularity of hardware ecosystems.

Software engineering is becoming relevant to hardware startups in two ways. Firstly, many hardware startups build their value propositions based on comprehensive systems of both software and hardware components, so-called hardware-related products. For instance, in a wearable device, the business value comes from not only the physical devices, but also the services of collecting, storing and analyzing personal data extracted from the devices. In such a context, development, operation and maintenance of hardware-related products involve software engineering processes and practices. Secondly, with the advancement in hardware prototyping and manufacturing, early-stage hardware engineering becomes more agile, similar to the agile software movement initiated almost a decade ago. Instead of a heavy design upfront, using tools and 3D printing allows shorter cycles of prototyping and more rigorous analysis of the product-market fit. Consequently, the approaches that are popular in software development, such as Agile, Lean Startup and Design Thinking, can be considered in hardware development and in general the production of the whole hardware-relate products.

Startups face with many challenges to survive in early stages, in which many are found to be related to engineering activities [3]. Every startup uses a certain approach to develop their product, and that, to some extent, has a direct impact on business objective and activities. For instance, products might need to be developed in a fast way to satisfy time-to-market demands (speed). Product development needs to be agile enough to support entrepreneurs in responding to sudden opportunities and threats when arriving (agility). While startups are considered a special context where traditional software engineering approaches might not be directly applicable, hardware startups are even more remarkable due to the special characteristics of the combination of hardware and software development. Although the research community is increasingly interested in product and business development paradigms in software startup contexts [4], empirical research on hardware startups is very limited.

We aim at investigating the characteristics of early-stage hardware-related product development and how they are aligned with business development. We intend to address the ability of startups to respond to changes from external environments. We conducted a multiple case study to provide insights on agility in hardware-related product development. The focus is on early-stage development activities, because later-stage startups might be easier to relate to existing hardware development approaches. Our research questions are:

> RQ1: What do agile mean to hardware startups?
> RQ2: How do hardware startups achieve agility in early stage product development?

The study is organized as follows: Section 2 presents a background about hardware startups, agility in startups and hardware development. Section 3 describes our research methodology, Section 4 presents our preliminary findings, and Section 5 contains discussions and future work.

## 2 RELATED WORK

The emergence of agile methods was a response to the inability of heavyweight, waterfall-like development methodologies to equip product development the responsiveness to change [5]. According to the Agile Manifesto, agile development values individuals and interactions over processes and tools, working software over comprehensive documentation, customer collaboration over contract negotiation, and responding to change over following a plan [6]. Abrahamsson et al. reviewed the literature in software development and clarified that agile development focusing on simplicity and speed, teamwork, customers and especially, working code [7]. In a general sense, agility can be defined as "*the capability to react and adapt to expected and unexpected changes within a dynamic environment constantly and quickly; and to use those changes (if possible) as an advantage*" [8].

Agile adoption in software startups is common. Nguyen-Duc et al. [9] reported that four out of five startups they studied have adopted agile development processes. Pantiuchina et al. surveyed 1526 software startups and found that speed related agile practices are used to a greater extent in comparison to quality practices [10]. Giardino et al. [4] observe that, to quickly validate the product in the market, software startups tend to use agile methods, but in an ad-hoc manner. These findings are reported in the context of software startups without any explicit investigation of hardware development.

Despite of the popularity of agile software development, the study of value and practices of agile in hardware development is not yet established [11]. Kaisti et al. suggested that Agile practices could be used in the embedded domain, but the practices need to be adapted to fit to the more constrained field of embedded product development [12]. Ronkainen et al. framed the development of embedded systems as 'hardware-related' software development [13]. Greene reported a positive experience of applying Agile approaches in firmware development in Intel [14]. Gustavsson reported a positive experience in first time adopting Agile approaches in hardware development in Erisson [15]. While these studies infer potential benefits of adopting Agile in hardware development, the investigated context is established companies or cooperates. Our work explores startup contexts, which presents a distinct environment from large companies or SMEs, due to their lack of necessary resources, temporal and evolving organizations and multiple influences [17].

## 3 RESEARCH APPROACH

### 3.1 Data collection and analysis

Given the unexplored nature of agile hardware development in startup context, we conducted a multiple exploratory case study [16]. We selected startups that (1) currently work in teams of both business developers and product developers, (2) operate at least six months and have at least a running prototype, (3) develop either wearable devices, IoT applications or embedded systems. In total, we selected 20 hardware startups.

Case selection and data collection were done in two phases. Phase 1 aimed at collecting pilot case studies and hardware startup cases with the focus on early stage activities, such as prototyping, agile practices and business development. In this phase, the cases were selected conveniently from our professional networks. Startups come from various countries, i.e. Norway, Finland, and Pakistan. Phase 2 aimed at collecting hardware startup cases with the focus on their practices of rapid and agile development, prototyping, technical debt and quality assurance. The selected startups were mainly from Norway.

The empirical data was collected from semi-structured interviews with key people of the startups, such as CEO, CTO and chief software/ hardware engineers. The interview guidelines are slightly different between two phases due to the different study scopes. However, both of the guidelines cover topics such as (1) business development approaches (2) engineering approaches, (3) prototyping, (4) agile adoption and (4) current challenges and wishes. These ensure sufficient data to address our RQs. All interviews were recorded and transcribed. In Phase 2, all participants signed consent forms before participation. There were 24 interviews conducted in total. The length of the interviews varied from 30 to 75 minutes.

Due to the limited understanding about the topic, we adopted an inductive approach in order to generate new knowledge. We applied a thematic synthesis process which is common for qualitative research in Software Engineering [2]. We started with reading through interview transcripts, identified relevant segment of texts and labeled them with codes. Codes were merged into themes, which later grouped into higher-order themes. The hierarchies of themes are presented by the thematic maps answering the RQs. For RQ1, we asked "Do you apply agile development in your company, i.e. Srum, Kanban, etc ?" and follow-up questions that reveals how entrepreneurs understand about Agile and its value. For RQ2, we asked in general the capacity of startups to be fast in the market, adaptive and react to changes. The qualitative maps are mainly from the data collected in Phase 1 of the research.

## 3.2 Case description

The whole sample includes 20 hardware startups from Norway (60%), Finland (15%), Pakistan (15%), Netherland (5%) and Italy (5%). All of the startups developed or are developing hardware-related products, falling into four categories:

- Personal tracking devices: patient monitoring, muscle operation measure.
- Machine interaction devices: smart board, smart home solutions, wheel chair controller.
- Utilities: camera, and camera's accessories, interactive toys, noise cancelling.
- Industrial IoT application: aerospace utilities, aquaculture tracking systems, ship tracking.

As shown in Figure 2, the median years of operation is two years in our sample. The number of employees ranges from three to 85, with the median of seven. There are three startups (15%) currently in the scaling phase and the rest in early stages (85%).

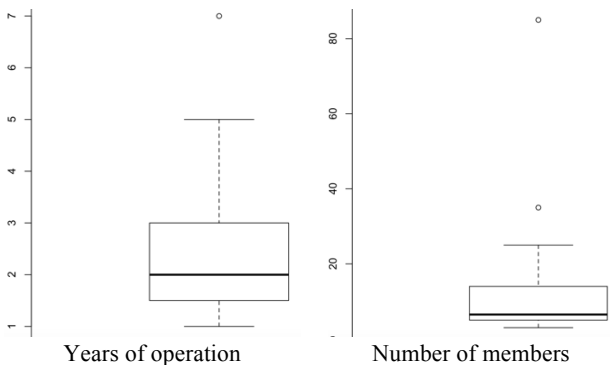

Figure 1: Startups' demographics

## 4 PRELIMINARY RESULTS

### 4.1. RQ1: What do agile mean to hardware startups?

Our preliminary analysis reveals that interviewees´ perceptions on agile development varied a lot. The concept of being agile is reflected in (1) principles, (2) practices and (3) scope, as shown in Figure 2.

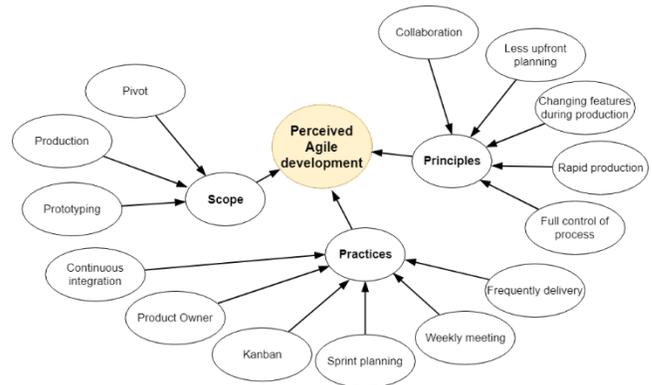

Figure 2: Thematic map of perceived agile development in hardware startups

Regarding the principles, entrepreneurs relate agile development with less upfront planning, short-term driven evolution of the startups. They also mentioned about the speed of prototyping, development and fast time-to-market when asked about agile. Entrepreneurs state that full controls of development activities and partnership will prepare themselves to respond to unexpected changes. Some startups also highlighted the importance of internal collaboration over defined processes.

Regarding the practices, entrepreneurs mention practices from different Agile frameworks, such as Scrum, XP and Kanban. There is no formal way of adopting Agile practices, but rather customized adoptions. Certain practices are mentioned by different interviewee, such as frequent delivery, sprint planning, Kanban and product owners. Some practices are mentioned by its tailored version, for instance, weekly meetings instead of daily standup meetings.

Regarding the scope, entrepreneurs relate agile to not only engineering activities, such as achieving rapid prototyping and agile product development, but also business level, with startup development and pivots.

Table 1 describes the extent that agile is adopted in hardware startups. We found that none of the investigated startups applies any Agile frameworks properly. More than half of the cases adopt some Agile practices, as mentioned in Table 1. Seven startups had considered about Agile and decided not to adopt it. There are two startups that do not know about Agile methodologies.

Table 1: The extent of agile practices adoption in hardware startups

| Level of agile adoption | No of cases |
|---|---|
| Have no idea about Agile methodologies | 2 |
| Know about Agile but do not use | 7 |
| Adopt some Agile practices | 11 |
| Strictly follow some Agile frameworks | 0 |

### 4.2. RQ2: How do hardware startups achieve agility in early stage product development?

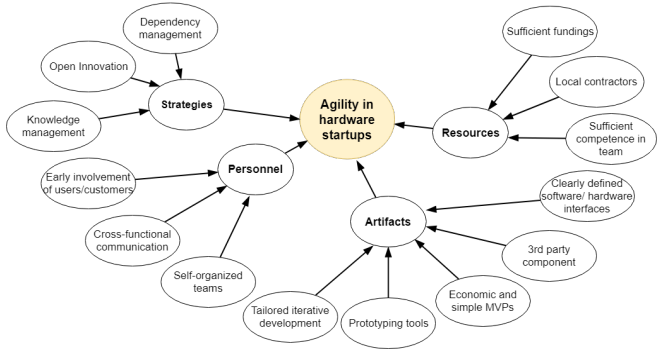

**Figure 3: Thematic map of enabling agility in early-stage hardware startups**

Our qualitative analysis reveals fourteen tactics to achieve agility in early-stage hardware startups, which are classified into four categories (1) strategic management, (2) personnel, (3) artifacts and (4) resources. Strategic management involves strategic thinking and decisions of CEOs that define the way business/ product are developed. Personnel includes interpersonal practices regarding to communication, coordination and collaboration, with internal and external stakeholders. Artifacts involve the practices of adopting and developing different types of artifacts, such as prototypes, tools, hardware/software components and interfaces. Resources refer to tactics that deal with using and managing financial/ human resources in early stage startups. We described two tactics from categories for illustrating the case as below:

**Tailored iterative development**: startups often build their own development approaches from known engineering practices, which are found suitable to their current working context. The commonalities among the startups is that they all adopted iterative development, i.e. having a regular releases. The releases are artifacts, i.e. a new prototype, or an increment of the previous prototype. A sprint duration varies among startups, from one week to four weeks. Since development of physical products usually takes longer time than implementation of software, the startups focused on defining measurable sub-goals that were part of a long-term plan. In our cases, throughout multiple Sprints, the whole physical product does not change much from the initial design. The adoption of certain Agile practices or approaches might be different between the development of hardware and software elements: "*Our electronics are relatively simple, while SW changes very much all the time. We are still trying to find what is the right way to do it*" (CTO of a Finnish startup)

**Prototyping tools and technologies**: hardware startups are beneficial from the advancement in prototyping technologies, such as 3D printing and CAD simulation tools. Almost all of our cases own or acquire 3D printing services, which enable them to make many physical prototypes in a short time. "*We solicit components, test different things and form factors, using a lot of 3D printing. We have invested in a better 3D printer so we can use the sensors on patients in the hospital. The cheaper printers make rough surfaces that will not be approved for medical use. This is a thing we otherwise would have to order from someone else, but that would take 3 weeks, so we removed it and invested in the better printer for our mechanical development." (CEO of a Norwegian startup)* It is also noted that even though 3D printing can be used for every physical component of a product, the printing takes time. In case of one case, it took 10 hours for a 3D printing of a small component. A lot of communication and changes happen already in the computer-based prototypes, i.e. with CAD designs.

## 5   CONCLUSIONS AND FUTURE WORK

Preliminary results reveal first observations about hardware startups. In 20 investigated startups, the concept of agility and being agile is known to mostly every case. Entrepreneurs have varied perceptions on agility, as it is not only about product development but also about reaction to business and pivot. We discovered some common Agile practices among hardware startups, such as iterative development, Kanban, and weekly meetings. We also found several tactics that hardware startups adopt to be agile in their business and product development.

The future direction would include further analysis of data in two directions. Firstly, we will explore in-depth the role of Agile value and practices in hardware startups. Given that no formal framework or process is adopted as it is, in which way Agile can be best beneficial to hardware development? Furthermore, we can map the challenges and practices of Agile with special characteristics of hardware startups. This brings following research questions:

- To what extent are Agile practices adopted in early stage hardware startups?
- What are the challenges when adopting Agile in startup hardware-related development?
- What Agile practices are beneficial in startup hardware development?

Secondly, agility is one dimension we investigated in hardware startups. Our second phase of data collection explores different topics, such as technical debts, quality assurance. It is especially interesting to explore if quality and speed can be both achieved, as they are both essential in hardware development. This brings following research questions:

- How are quality aspects assured in hardware startups?
- How is technical debt managed in hardware startups?
- How do hardware startups achieve balance between speed and quality?